# Impact of 2D-Graphene on SiN Passivated AlGaN/GaN MIS-HEMTs under Mist Exposure

M. Fátima Romero, Alberto Boscá, Jorge Pedrós, Javier Martínez, Rajveer Fandan, Tomás Palacios, *Fellow, IEEE*, and Fernando Calle, *Member, IEEE*

*Abstract*— The effect of a two dimensional (2D) graphene layer (GL) on top of the silicon nitride (SiN) passivation layer of AlGaN/GaN metal-insulator-semiconductor high-electron-mobility transistors (MIS-HEMTs) has been systematically analyzed. Results showed that in the devices without the GL, the maximum drain current density ($I_{D,max}$) and the maximum transconductance ($g_{m,max}$) decreased gradually as the mist exposure time increased, up to 23% and 10%, respectively. Moreover, the gate lag ratio (GLR) increased around 10% during mist exposure. In contrast, devices with a GL showed a robust behavior and not significant changes in the electrical characteristics in both DC and pulsed conditions. The origin of these behaviors has been discussed and the results pointed to the GL as the key factor for improving the moisture resistance of the SiN passivation layer.

## I. INTRODUCTION

GALLIUM nitride based high electron mobility transistors (HEMTs) are very promising devices for high power, high frequency and high temperature applications [1]–[4]. However, there are still some relevant issues, such as, high gate leakage current and current collapse, which limit the full potential performance of this kind of devices and their full market commercialization [5]–[7]. In fact, these problems lead to limited gate voltage swing, reduced breakdown voltage, increased static power dissipation, and decreased RF performance. Therefore, insulated-gate and surface-passivation structures are needed to mitigate these challenges. In particular, the use of as *in-situ* SiN layer on top of AlGaN/GaN HEMT structures has showed recently to be feasible and advantageous to reduced AlGaN relaxation, increased sheet carrier concentration ($n_s$), improved ohmic contacts and surface protection during processing [8], [9]. On the other hand, SiN passivation is widely used to prevent current collapse, even though the exact mechanism is still not clear. Gao *et al* recently proposed that water-related redox reactions play a significant role in the physical origin of surface trapping states both in unpassivated and passivated AlGaN/GaN HEMTs [10], [11]. Although they showed that a thick SiN passivation and a fluorocarbon top layer can mitigate this effect [12], a robust solution to avoid water adsorption and diffusion is still under research.

Graphene has been proved to be compatible with GaN technology, where graphene capping layers can improve the thermal management of AlGaN/GaN HEMTs [13]. Also, the thermal stability of GaN Schottky diodes at elevated temperatures (550 K) was improved by graphene likely acting as an impenetrable barrier to the diffusion of contaminants across the interface [14]. In this letter, we propose the use of graphene as two dimensional and hydrophobic material compatible with GaN-technology, in combination with the SiN layer passivation, to efficiently prevent the trapping effects, in particular those water-related, that affect AlGaN/GaN HEMT devices. Besides, the use of the graphene layer (GL) allows to use a thinner SiN layer, therefore reducing the fringing capacitance [15] without compromising the water-related current collapse effects.

This work was supported in part by the RUE (CSD2009-00046), CAVE (TEC2012-38247) and GRAFAGEN (ENE2013-47904-C3) projects, funded by Ministerio de Ciencia e Innovación and Ministerio de Economía y Competitividad (MINECO) of Spain. This work has also received funding from the European Union's Horizon 2020 research and innovation programme under grant agreement No 642688. M. F. R. acknowledges funding from MINECO Juan de la Cierva-Incorporación grant (IJCI-2014-19473). J. P. acknowledges MINECO Ramón y Cajal grant (RYC-2015-18968). T. P. would like to thank the ONR PECASE and the AFOSR FATE MURI for partial support.

M. F. Romero, A. Boscá, J. Pedrós, J. Martínez, R. Fandan, and F. Calle are with the Instituto de Sistemas Optoelectrónicos y Microtecnología and the Departamento de Ingeniería Electrónica, ETSI de Telecomunicación, Universidad Politécnica de Madrid, 28040 Madrid, Spain (e-mail: fromero@isom.upm.es).

T. Palacios is with the Department of Electrical Engineering and Computer Science, Massachusetts Institute of Technology, Cambridge, Massachusetts 02139, USA.

## II. EXPERIMENTAL PROCEDURE

The devices were fabricated on AlGaN/GaN heterostructures (HS) grown on (111) silicon wafer with an *in situ* SiN cap layer grown by metal-organic chemical vapor deposition. The details of the HS are SiN (3 nm)/Al$_{0.25}$Ga$_{0.75}$N (25 nm) / GaN (2 µm). The sheet resistance measured by Lehighton contactless measurements is 476 Ω/sq. 100-nm-deep device isolation was carried out by inductive coupling etching (ICP) using Cl$_2$/Ar-based process. Prior to the formation of the ohmic contacts the SiN layer was selectively removed, by reactive ion etching (RIE), using SF$_6$ plasma. Before metal deposition, a clean low-power O$_2$-plasma descum and dilute HCl were used. The ohmic contact metallization was formed by Ti/Al/Ni/Au (20/120/40/80 nm) deposited using e-beam evaporation and was rapid-thermal-annealed for 30 s at 850°C. The contact resistance ($R_c$= 0.54 Ω·mm) and sheet resistance ($R_{sheet}$= 452 Ω/sq) were calculated using the standard TLM method. The gate metallization was Ni/Au (20/200 nm) on top of the cap layer. Then, a passivation layer of $SiN_x$ (100 nm) by plasma enhanced chemical vapor deposition (PECVD) was deposited. Finally, the $SiN_x$ was patterned and etched from the regions of the ohmic contacts, and a Ti/Au (20/200 nm) overlay metal was deposited.

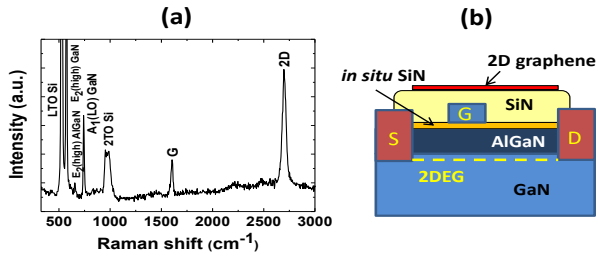

Fig. 1. (a) Raman spectrum of the CVD 2D graphene layer transferred onto the *in situ* SiN/AlGaN/GaN HEMT devices, and (b) schematic cross-section of the devices with graphene.

The wafer was divided into two pieces, one piece (named A) was kept as reference and on the other piece (named B) a film of monolayer graphene grown by chemical vapor deposition was transferred from a Cu substrate [16]. In this case, to have a good adhesion to the surface, the sample was vacuum dried for 24 hours, in order to remove the water layer between the GL and the substrate surface during the graphene transfer. The Raman spectrum in Fig. 1(a) shows a negligible D peak and a ratio of 2D peak intensity to G peak intensity I(2D)/I(G) > 2, which is consistent with defect-free single layer graphene [17]. After the transfer, the graphene was selectively etched away from the gate, drain and source contacts to avoid short circuits between these electrodes, by means of $O_2$ plasma using ICP (20 W, for 30 s at room temperature (RT)).

The layout of the metal-insulator-semiconductor HEMT (MIS-HEMT) consisted of one finger devices with gate length $L_G$= 4 and 5 µm, gate width $W_G$ = 100 µm, and gate to drain distance $L_{GD}$ = 15 µm. A schematic of the cross section of the MIS-HEMT device capped with a GL is shown in Fig. 1(b). Electrical characterization using DC and pulsed current-voltage (I-V) measurements was carried out at RT in devices with and without the GL before, during and after being exposed to mist focused directly to the HEMT sample from a jet nebulizer system (Fig. 2).

## III. RESULTS AND DISCUSSION

Firstly, the effect of adding a GL on top of the SiN passivation layer was assessed, so sample B was tested before and after the graphene transfer. The maximum drain current density ($I_{D,max}$= 0.3 A/mm at $V_{GS}$= 0 V), the maximum transconductance ($g_{m,max}$= 115 mS/mm), the threshold voltage ($V_{th}$= -3.5 V), the gate leakage current at $V_{GS}$= -6 V and $V_{DS}$= 15 V ($I_{G,leak}$= $10^{-6}$ A/mm) and the off-state drain current $V_{GS}$= -6 V and $V_{DS}$= 15 V ($I_{D,off}$= $10^{-6}$ – $10^{-5}$ A/mm) did not change, as shown in Fig. 3. The rough values in brackets correspond to the $L_G$= 5 µm devices, but similar results were obtained for those with $L_G$= 4 µm, and they are in good agreement with the literature [8]. The results above are consistent with what is expected considering that the GL is not in contact with the gate neither the drain nor the source metals, so no bias voltage is applied to the GL.

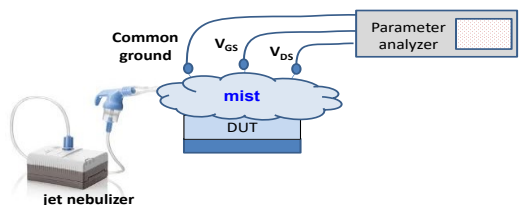

Fig. 2. Schematic of the experimental setup used for characterizing the device under test (DUT) during the mist exposure.

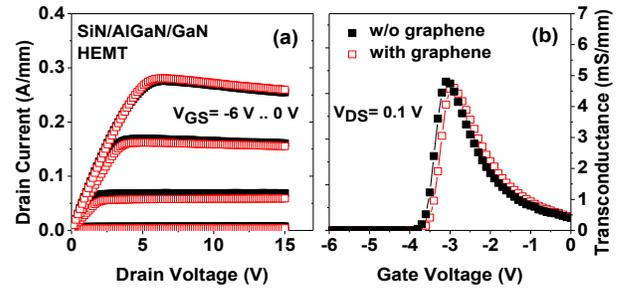

Fig. 3. (a) Output and (b) transfer characteristics of the MIS-HEMTs before and after the transfer of the GL. $L_G$= 5 µm and $L_{GD}$ = 15 µm.

Both A and B samples were characterized before, during and after the mist exposure. Figure 4 provides direct comparison of the normalized output characteristics of both kinds of devices (a) without and (b) with the GL. The normalization was done respect to the $I_{DS,max}$ (at $V_{GS}$= 0 V) before the mist exposure. In the case of sample A, the $I_{D,max}$ decreased gradually as the mist exposure time increased until it reached a minimum value after about 10 min, which could be a transient of the possible water permeation process. This reduction ranges from 12 to 23%, depending on the device, though no correlation was observed with the device geometry. Interestingly, it was observed that the higher the negative gate voltage, the larger the change of $I_D$ caused by mist. A possible explanation is that more negative voltages at the gate might accelerate the process of permeation of the water molecules through the SiN layer, and thus affect the AlGaN surface.

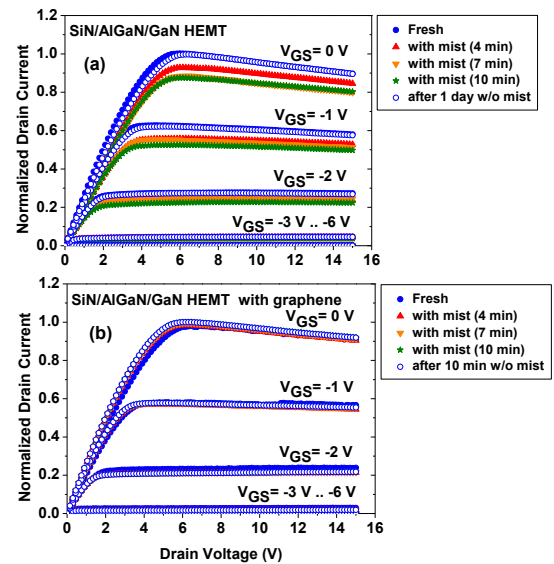

Fig. 4. Normalized $I_D$-$V_{DS}$ ($V_{GS}$) characteristics before, during and after the mist exposure to the MIS-HEMTs (a) without (sample A) and (b) with a top GL (sample B). $L_G$= 5 µm and $L_{GD}$ = 15 µm.

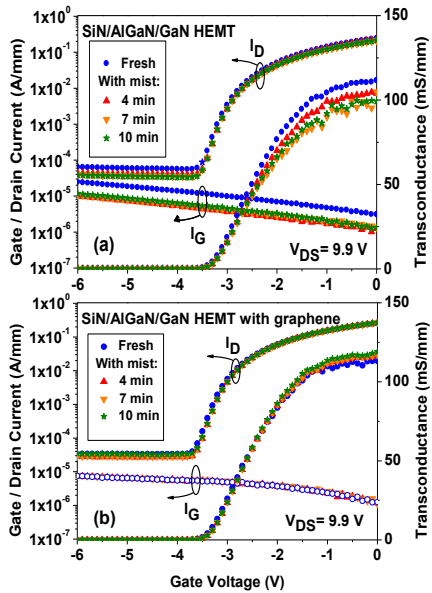

Fig. 5. $I_D$- and $I_G$-$V_{GS}$ ($V_{DS}$) characteristics before and during the mist exposure to the MIS-HEMTs (a) without (sample A) and (b) with a top GL (sample B). $L_G$= 5 µm and $L_{GD}$ = 15 µm.

However, further experiments should be done to confirm this hypothesis. In contrast, the devices in sample B did not show significant changes in the $I_{D,max}$. It is noteworthy that the changes in the output characteristics in sample A are reversible after the mist exposure is stopped. The $I_{D,max}$ values go back slowly to the fresh values after about 1 day resting, as shown in Fig. 4 (a). About 10 devices were tested under mist conditions. In addition, the test was repeated after several days in some of the devices, with and without graphene, and the results were reproducible.

Regarding the transfer characteristics, a similar behavior was obtained for both kinds of devices, featured by a slight decrease (~10%) in $g_{m,max}$ for the MIS-HEMT devices in sample A (Fig. 5 (a)), and hardly no changes in the devices in sample B (Fig. 5(b)). Interestingly, the $V_{th}$ and the leakage currents ($I_{G,leak}$ and $I_{D,off}$) did not change significantly during the mist exposure in neither sample A nor B. These results reveal that the *in situ* SiN layer keeps a robust gate-to-channel control and the mist does not affect the 2DEG underneath the gate in either type of sample. However, the reduction of $I_{D,max}$ in MIS-HEMT devices without GL suggests that the mist exposure is able to affect the 2DEG in the region between the gate and the drain contacts, likely due to the water molecules that may permeate through the SiN layer and affect the AlGaN surface [18]. In fact, Mehandru *et al* showed that bonding of polar liquid molecules, such as water, appear to alter the polarization-induced positive AlGaN surface charge, leading to changes in the channel carrier density and hence the drain-source current [19].

In order to analyze further the effect of the moisture on the AlGaN surface and in the associated trapping phenomena, pulsed I-V measurements were also carried out. In order to minimize the changes in the drain current due to the mist exposure time, the pulsed measurement were carried out just after the first 10 min of exposure, once the drain current drop is stable. The gate lag ratio (GLR) was defined as the ratio of the pulsed $I_D$ with quiescent points ($V_{GSQ}$, $V_{DSQ}$) = (-6 V, 0 V) and (0 V, 0 V), in order to avoid self-heating effects. Results show that the GLR is ~ 1 for both fresh samples, indicating almost no current collapse. This is explained by the fact that the *in situ* SiN layer is protecting effectively the AlGaN surface, in good agreement with the literature [8], [9]. However, in the presence of mist, the GLR of the devices with the GL keeps stable around 1, while the devices without graphene suffer from around 10% higher gate lag respect to the fresh devices.

Hence, these results confirm that the moisture plays a critical role in the degradation of the characteristics of conventional GaN-based MIS-HEMT devices, in agreement with recent reports indicating that the moisture on the AlGaN surface can modify significantly the 2DEG [11], [20]. Moreover, Gao *et al* showed that this adverse effects are also present even with a 20-nm-thick SiN passivation layer, but it can be mitigated with a thicker SiN layer (200 nm) [11]. In the present case, the passivation layer is 100 nm thick, and according to the results it seems not to be thick enough to impede either the water permeation through the SiN layer and reaching the AlGaN surface [17], or the electrons from the gate in reverse bias being injected onto the passivation surface, as Gao *et al* suggested [11]. Therefore, the water molecules either adsorbed on top of the SiN layer or permeated through it reaching the AlGaN surface could act as surface trapping centers. Therefore, these trapped electrons could form a second gate depleting the channel in the case of sample A.

On the other hand, the devices with a top GL (sample B) show the beneficial effects of the graphene which is able to repel effectively the moisture due to its hydrophobic nature, making the device impermeable, thus avoiding the water-related trapping effects that might affect the AlGaN surface during the mist exposure. Although the experimental set-up could be improved using a chamber with a better atmosphere control, these results present direct evidence of the improvement in the AlGaN/GaN MIS-HEMT device performance under harsh atmospheres when a protective GL is placed on top of the SiN passivation.

## IV. CONCLUSION

The effects of a GL on top of the SiN passivation layer of an AlGaN/GaN HEMT with an *in situ* SiN cap layer have been assessed before, during and after mist exposure. A gradual degradation in the $I_{D,max}$ and $g_{m,max}$ values was observed in the devices without a GL as the mist exposure time increased. In contrast, the characteristics of the devices with the GL were stable during the process, even under pulsed conditions. The results highlight the critical role of the GL in improving the resistance against moisture of the SiN passivation layer by preserving the electrical characteristics of the AlGaN/GaN MIS-HEMTs, mainly due to the excellent hydrophobic property of graphene.

## ACKNOWLEDGMENT

The authors thank A. del Campo (ICV-CSIC) for his assistance with the Raman measurements. M. J. Tadjer, A. D. Koehler and T. J. Anderson are acknowledged for their technical assistance.

## References

[1] U. K. Mishra, P. Parikh, and Y.-F. Wu, "AlGaN/GaN HEMTs—an overview of device operations and applications," Proc. IEEE, vol. 90, pp. 1022-1031, Jun. 2002. DOI: 10.1109/JPROC.2002.1021567

[2] K. Shinohara, D. C. Regan, Y. Tang, A. L. Corrion, D. F. Brown, J. C. Wong, J. F. Robinson, H. H. Fung, A. Schmitz, T. C. Oh, S. J. Kim, P. S. Chen, R. G. Nagele, A. D. Margomenos, and M. Micovic. "Scaling of GaN HEMTs and Schottky Diodes for Submillimeter-Wave MMIC Applications". IEEE Trans. Electron Devices, vol. 60, pp. 2982-2996, Oct. 2013. DOI: 10.1109/TED.2013.2268160

[3] M. Van Hove, X. Kang, S. Stoffels, D. Wellekens, N. Ronchi, R. Venegas, K. Geens, and S. Decoutere, "Fabrication and performance of Au-Free AlGaN/GaN-on-Silicon power devices with and gate dielectrics". IEEE Trans. Electron Devices, vol. 60, pp. 3071-3078, Oct. 2013. DOI: 10.1109/TED.2013.2274730

[4] A. Tarakji, X. Hu, A. Koudymov, G. Simin, J. Yang, M. A. Khan, M. S. Shur, R. Gaska, "DC and microwave performance of a GaN/AlGaN MOSHFET under high temperature stress", Solid-State Electronics, vol. 46, pp. 1211–1214, Aug. 2002. DOI: 10.1016/S0038-1101(02)00015-1

[5] D. Yan, H. Lu, D. Cao, D. Chen, R. Zhang, and Y. Zheng, "On the reverse gate leakage current of AlGaN/GaN high electron mobility transistors", Appl. Phys. Lett., vol. 97, pp. 153503 (1-3), Sep. 2010. DOI: http://dx.doi.org/10.1063/1.3499364

[6] G. Meneghesso, G. Verzellesi, F. Danesin, F. Rampazzo, F. Zanon, A.Tazzoli, M. Meneghini, and E. Zanoni. "Reliability of GaN high-electron-mobility transistors: state of the art and perspectives". IEEE Trans Dev Mater Reliab 8, pp. 332-343, Jun. 2008. DOI: 10.1109/TDMR.2008.923743

[7] M. Dammann, W. Pletschen, P. Waltereit, W. Bronner, R. Quay, S. Müller, M. Mikulla, O. Ambacher, P.J. van der Wel, S. Murad, T. Rödle, R. Behtash, F. Bourgeois, K. Riepe, M. Fagerlind, E.Ö. Sveinbjörnsson. "Reliability and degradation mechanism of AlGaN/GaN HEMTs for next generation mobile communication systems". Microelectronics Reliability, vol. 49, pp. 474–477, May 2009. DOI: 10.1016/j.microrel.2009.02.005

[8] M. Germain, K. Cheng, J. Derluyn, S. Degroote, J. Das, A. Lorenz, D. Marcon, M. Van Hove, M. Leys, and G. Borghs. "In-situ passivation combined with GaN buffer optimization for extremely low current dispersion and low gate leakage in $Si_3N_4$/AlGaN/GaN HEMT devices on Si (111)". Phys. stat. sol. c, vol. 5, pp. 2010–2012, Apr. 2008. DOI: 10.1002/pssc.200778688

[9] H. Jiang, C. Liu, Y. Chen, X. Lu, C. W. Tang, K. M. Lau. "Investigation of in situ SiN as gate dielectric and surface passivation for GaN MISHEMTs". IEEE Trans. Electron Devices, vol. 99, pp. 1-8, Mar. 2017. DOI: 10.1109/TED.2016.2638855

[10] F. Gao, D. Chen, B. Lu, H. L. Tuller, C. V. Thompson, S. Keller, U. K. Mishra, and T. Palacios. "Impact of Moisture and Fluorocarbon Passivation on the Current Collapse of AlGaN/GaN HEMTs", IEEE Electr. Device Letters, vol. 33, pp. 1378-1380, Oct. 2012. DOI: 10.1109/LED.2012.2206556

[11] F. Gao, D. Chen, H. L. Tuller, C. V. Thompson, and T. Palacios."On the redox origin of surface trapping in AlGaN/GaN high electron mobility transistors". J. Appl. Phys., vol. 115, pp. 124506, Mar. 2014. DOI: http://dx.doi.org/10.1063/1.4869738

[12] F. Gao. "Impact of Electrochemical Process on the Degradation Mechanisms of AlGaN/GaN HEMTs", Ph.D. dissertation. Dep. Mat. Sci. and Eng., Massachusetts Inst. Tech. (MIT), Cambridge, MA, USA, 2014.

[13] Z. Yan, G. Liu, J. M. Khan and A. Balandin. "Graphene quilts for thermal management of high-power GaN transistors". Nature Communications vol. 3, pp. 827 (1-8), May 2012. DOI: 10.1038/ncomms1828

[14] S. Tongay, M. Lemaitre, T. Schumann, K. Berke, B. R. Appleton, B. Gila, and A. F. Hebard. "Graphene/GaN Schottky diodes: Stability at elevated temperatures". Appl. Phys. Lett. vol. 99, pp. 102102 (1-3), Sept. 2011. DOI: 10.1063/1.3628315

[15] D. S. Lee, O. Laboutin, Y. Cao, W. Johnson, E. Beam, A. Ketterson, M. Schuette, P. Saunier, T. Palacios. "Impact of $Al_2O_3$ Passivation Thickness in Highly Scaled GaN HEMTs". IEEE Electron Device Letters 33, pp. 976 – 978 (2012). DOI: 10.1109/LED.2012.2194691

[16] A. Boscá, J. Pedrós, J. Martínez, T. Palacios, and F. Calle. "Automatic graphene transfer system for improved material quality and efficiency", Sci. Rep., vol. 6, pp. 21676 (1-8), Feb.2016. DOI: 10.1038/srep21676

[17] X. Li, W. Cai, J. An, S. Kim, J. Nah, D. Yang, R. Piner, A. Velamakanni, I. Jung, E. Tutuc, S. K. Banerjee, L. Colombo, R. S. Ruoff. "Large-area synthesis of high-quality and uniform graphene films on copper foils". Science vol. 324, pp. 1312-1314, Jun. 2009. DOI: 10.1126/science.1171245

[18] D.S. Wuu, W.C. Lo, C.C. Chiang, H.B. Lin, L.S. Chang, R.H. Horng, C.L. Huang, Y.J. Gao."Water and oxygen permeation of silicon nitride films prepared by plasma-enhanced chemical vapor deposition", Surface & Coatings Technology, vol. 198, pp. 114– 117, Aug. 2005. DOI: https://doi.org/10.1016/j.surfcoat.2004.10.034

[19] R. Mehandru, B. Luo, B.S. Kang, J. Kim, F. Ren, S.J. Pearton, C.-C. Pan, G.-T. Chen, J.-I. Chyi. "AlGaN/GaN HEMT based liquid sensors", Solid-State Electronics, vol. 48, pp. 351–353, 2004. DOI: 10.1016/S0038-1101(03)00318-6

[20] S. Ozaki, K. Makiyama, T. Ohki, Y. Kamada, M. Sato, Y. Niida, N. Okamoto, and K. Joshin. "Millimeter-Wave GaN HEMTs with cavity-gate structure using MSQ-Based inter-layer dielectric", IEEE Trans. Semicond. Manufacturing, vol. 29, pp. 370-375, Nov. 2016. DOI: 10.1109/TSM.2016.2599184